\documentstyle[12pt,epsfig]{article}
\begin{document}
\setlength{\headheight}{0in}
\setlength{\headsep}{0in}
\setlength{\topskip}{1ex}
\setlength{\textheight}{8.5in}
\setlength{\topmargin}{0.5cm}
\setlength{\baselineskip}{0.24in}
\catcode`@=11
\long\def\@caption#1[#2]#3{\par\addcontentsline{\csname
  ext@#1\endcsname}{#1}{\protect\numberline{\csname
  the#1\endcsname}{\ignorespaces #2}}\begingroup
    \small
    \@parboxrestore
    \@makecaption{\csname fnum@#1\endcsname}{\ignorespaces #3}\par
  \endgroup}
\catcode`@=12
\def\slashchar#1{\setbox0=\hbox{$#1$}           
   \dimen0=\wd0                                 
   \setbox1=\hbox{/} \dimen1=\wd1               
   \ifdim\dimen0>\dimen1                        
      \rlap{\hbox to \dimen0{\hfil/\hfil}}      
      #1                                        
   \else                                        
      \rlap{\hbox to \dimen1{\hfil$#1$\hfil}}   
      /                                         
   \fi}                                         %
\newcommand{\newc}{\newcommand}
\def\be{\begin{equation}}
\def\ee{\end{equation}}
\def\bea{\begin{eqnarray}}
\def\eea{\end{eqnarray}}
\def\simlt{\stackrel{<}{{}_\sim}}
\def\simgt{\stackrel{>}{{}_\sim}}
\begin{titlepage}
\begin{flushright}
{\setlength{\baselineskip}{0.18in}
{\normalsize
IC/00/46\\
UCRHEP-T272\\
hep-ph/0004148\\
April 2000\\
}}
\end{flushright}
\vskip 2cm
\begin{center}

{\Large\bf 
Relaxation of the Dynamical Gluino Phase\\
and Unambiguous Electric Dipole Moments\\}

\vskip 2.5cm

{\large
D. A. Demir$^1$ and Ernest Ma$^2$\\}

\vskip 0.5cm
{\setlength{\baselineskip}{0.18in}
{\normalsize $^1$ The Abdus Salam International Centre for Theoretical 
Physics,\\ I-34100 Trieste, Italy\\[5pt]
$^2$ Department of Physics, University of California,\\ 
Riverside, California 92521, USA\\}}

\end{center}
\vskip 2.5cm
\begin{abstract}
We propose a new axionic solution of the strong CP problem with a Peccei-Quinn 
mechanism using the gluino rather than quarks.  The spontaneous breaking of 
this new global U(1) at 10$^{11}~{\rm GeV}$ also generates the supersymmetry 
breaking scale of 1 $\rm{TeV}$ (solving the so-called $\mu$ problem at the same 
time) and results in the MSSM (Minimal Supersymmetric 
Standard Model) with R parity conservation.  In this framework, electric 
dipole moments become calculable without ambiguity.
\end{abstract}
\end{titlepage}

\setcounter{footnote}{0}
\setcounter{page}{1}
\setcounter{section}{0}
\setcounter{subsection}{0}
\setcounter{subsubsection}{0}

CP nonconservation is a fundamental issue in particle physics.  We know 
that CP is not conserved in $K - \overline K$ mixing (i.e.~$\epsilon \neq 
0$)\cite{exp1} and in $K$ decay (i.e.~$\epsilon' \neq 0$)\cite{exp2}.  
However, only an upper limit ($0.63 \times 10^{-25} e\cdot$cm) exists for 
the electric dipole moment ($edm$) of the neutron\cite{exp3}.  This may not 
be so bothersome until we realize that the currently accepted theory of strong 
interactions, i.e.~quantum chromodynamics (QCD), actually violates CP 
through the instanton-induced term\cite{theta}
\begin{equation}
{\cal L}_{\theta} = \theta_{QCD} {g_s^2 \over 64 \pi^2} \epsilon_{\mu \nu 
\alpha \beta} G_a^{\mu \nu} G_a^{\alpha \beta},
\end{equation}
where $g_s$ is the strong coupling constant, and
\begin{equation}
G_a^{\mu \nu} = \partial^\mu G_a^\nu - \partial^\nu G_a^\mu + g_s f_{abc} 
G_b^\mu G_c^\nu
\end{equation}
is the gluonic field strength.  The value of $\theta_{QCD}$ must then be 
less than $10^{-10}$ in magnitude (instead of the expected order of unity) 
to account for\cite{edm} the nonobservation of the neutron $edm$.  This is 
known as the strong CP problem.

Another fundamental issue in particle physics is supersymmetry.  It allows 
us to solve the hierarchy problem so that our effective theory at the 
electroweak energy scale ($M_W$) is protected against large radiative 
corrections.  However, this requires the scale of soft supersymmetry 
breaking ($M_{SUSY}$) to be not much higher than $M_W$.  There is no  
theoretical understanding of why the two scales {\it must} be related in 
this way.

In the following we address the question of how the strong CP problem is to be 
solved in the context of supersymmetry.  We find that it can be achieved 
with a new kind of axion\cite{pq,ww} which couples to the gluino rather than 
to quarks.  In a natural implementation of this idea, we find that the 
breaking of supersymmetry must have the same origin as the axion.  The scale 
of electroweak symmetry breaking is also related.  This works because the 
so-called $\mu$ problem is being solved along the way.

In our framework, the $\theta_{QCD}$ contribution to any quark $edm$ is 
canceled exactly by the minimization of the dynamical gluino phase.  Hence 
the calculation of $edm$'s in the MSSM (Minimal Supersymmetric Standard 
Model) becomes unambiguous.  The possibility of cancellation\cite{cancel} 
among other different CP nonconserving contributions\cite{susycp} to $edm$'s 
can now be pursued without fear of contradiction. 

With the addition of colored fermions, the parameter $\theta_{QCD}$ of Eq.~(1) 
is replaced by
\begin{equation}
\overline \theta = \theta_{QCD} - Arg ~Det ~M_u M_d - 3 ~Arg ~M_{\tilde g},
\end{equation}
where $M_u$ and $M_d$ are the respective mass matrices of the charge = 2/3 
and charge = --1/3 quarks, and 
$M_{\tilde g}$ is the mass of the gluino.  The famous Peccei-Quinn 
solution\cite{pq} is to introduce a dynamical phase to the quark masses 
which then relaxes $\overline \theta$ to zero.  The specific realization of 
this requires an axion\cite{ww} which is ruled out experimentally\cite{exp4}. 
Two other axionic solutions have been proposed which are at present 
consistent with all observations.  The DFSZ solution\cite{dfsz} introduces 
a singlet scalar field as the source of the axion but its mixing with the 
doublet scalar fields which couple to the quarks is very much suppressed. 
The KSVZ solution\cite{ksvz} introduces new heavy quarks so that 
the axion does not even couple directly to the usual quarks.  Neither scheme 
requires supersymmetry.

In the context of supersymmetry however, it is clear that the simplest 
and {\it most natural} thing to do  is to attach the axion to the gluino 
rather than to the quarks in Eq.~(3).  Because of the structure of 
supersymmetry, all other  superparticles will be similarly affected.  
This is then a very strong hint that 
it may have something to do with the breaking of supersymmetry.  As shown 
below with our proposed singlet complex scalar field $S$, whose phase 
contains the axion, all soft supersymmetry breaking parameters are of 
order $|\langle S \rangle|^2/M_{Pl}$, where $M_{Pl} \sim 10^{19}$ GeV is 
the Planck mass.  Hence a value of $10^{11}$ GeV for $|\langle S \rangle|$, 
which is allowed by astrophysical and cosmological constraints, would imply 
$M_{SUSY} \sim 1$ TeV.  Since the electroweak symmetry breaking terms are 
also among this group, it does not require any stretch of the imagination to 
find $M_W$ and $M_{SUSY}$ to be only an order of magnitude apart.

It is known\cite{dith} that a continuous global $U(1)_R$ symmetry\cite{transf} 
can be defined for the MSSM.  The quark ($\hat Q, \hat u^c, \hat d^c$) and 
lepton ($\hat L, \hat e^c$) chiral superfields have $R = +1$ whereas the Higgs 
($\hat H_u, \hat H_d$) chiral superfields and the vector superfields have 
$R = 0$.  The superpotential
\begin{eqnarray}
\hat{W} = \mu \hat{H}_{u} \cdot \hat{H}_{d} + h_{u} \hat{Q} \cdot \hat{H}_{u} 
~\hat u^c + h_{d} \hat{Q} \cdot \hat{H}_{d} ~\hat d^c + h_{e} \hat{L}\cdot 
\hat{H}_{d} ~\hat e^c
\end{eqnarray}
has $R = +2$ except for the $\mu$ term (which has $R = 0$).  In the above, 
the Yukawa couplings $h_{u,d,e}$ are nonhermitian matrices in  
flavor space. The resulting Lagrangian is then invariant only with respect 
to the usual discrete $R$ parity, i.e.
\begin{equation}
R \equiv (-1)^{3 B + L + 2J},
\end{equation}
where $B$ is baryon number, $L$ lepton number, and $J$ spin angular momentum, 
hence $R$ is even for particles and odd for superparticles.

We now propose to make $U(1)_R$ an exact global symmetry of the supersymmetric 
Lagrangian, as well as that of all the supersymmetric breaking terms. 
We introduce the composite operator 
\begin{eqnarray}
\label{composite}
\mu(\hat{S})\equiv \frac{1}{M_{Pl}} \left(\hat{S}\right)^{2}~,
\end{eqnarray} 
where the singlet superfield $\hat S$ has $R = +1$.  Our model is then defined 
by the new superpotential
\begin{eqnarray}
\label{superpot}
\hat{W}&=& \mu(\hat{S})~\hat{H}_{u}\cdot \hat{H}_{d}+ 
m_s^2 \mu(\hat S) \nonumber \\ &+&h_{u} \hat{Q}\cdot 
\hat{H}_{u} ~\hat u^c +h_{d} \hat{Q}\cdot \hat{H}_{d} ~\hat d^c + h_{e} 
\hat{L}\cdot \hat{H}_{d} ~\hat e^c,
\end{eqnarray}
which has $R = +2$, thus yielding a supersymmetric Lagrangian which is 
invariant under $U(1)_R$, together with the following set of 
supersymmetry breaking terms which are also invariant under $U(1)_R$:
\begin{eqnarray}
\label{soft}
\Delta{\cal{L}}&=&|\mu(S)|^{2}\Big[\tilde{Q}^{\dagger} Y_{Q} \tilde{Q} + 
\tilde{u^c}^{\dagger} Y_u \tilde{u^c}+\tilde{d^c}^{\dagger} Y_d
\tilde{d^c}+\tilde{L}^{\dagger} Y_{L} \tilde{L}+\tilde{e^c}^{\dagger} Y_e 
\tilde{e^c}\Big]\nonumber\\&+& \Big\{\mu(S)^{\dagger}\big[k_{u} \tilde{Q}
\cdot {H}_{u}~\tilde{u^c}+k_{d} \tilde{Q}\cdot {H}_{d} ~\tilde{d^c} + k_{e} 
\tilde{L}\cdot {H}_{d} ~\tilde{e^c}\big]+ h. c.\Big\}\nonumber\\ &+& 
|\mu(S)|^{2}\big[y_{u} |H_u|^{2}+y_{d} |H_d|^{2}+\left(k_{\mu}H_{u}\cdot 
H_{d} + h. c. \right)\big]\nonumber\\ &+&\Big\{\mu(S)^{\dagger}\big[k_{3} 
\tilde{\lambda}^{a}_{3}\tilde{\lambda}^{a}_{3} + k_2 \tilde{\lambda}^{i}_{2}
\tilde{\lambda}^{i}_{2}+ k_1 \tilde{\lambda}_{1}\tilde{\lambda}_{1}\big]+ 
h. c. \Big\},
\end{eqnarray}
where $\tilde \lambda_3^a$ is the gluino octet, $\tilde \lambda_2^i$ the 
$SU(2)_L$ gaugino triplet, and $\tilde \lambda_1$ the $U(1)_Y$ gaugino 
singlet.  The parameters $k_{1,2,3}$ and $k_\mu$ are complex, whereas 
$y_{u,d}$ are real.  The matrices $Y_{Q,L}$ and $Y_{u,d,e}$ are hermitian, 
whereas $k_{u,d,e}$ are nonhermitian.  Obviously, we have assumed in the 
above that the source of all supersymmetry breaking terms is $\mu(S)$.  
Together with $U(1)_R$, this solves the so-called $\mu$ problem in the MSSM, 
because the scale of $\mu(S)$ is $|\langle S \rangle|^2/M_{Pl}$ which is of 
order 1 TeV for $|\langle S \rangle| \sim 10^{11}$ GeV, instead of the 
typical grand unification scale of $10^{16}$ GeV.

The Lagrangian ${\Delta \cal{L}}$ describes the interaction of $\mu(S)$ with 
sfermions, Higgs doublets and gauginos only.  However, a complete description 
of our model requires the self interactions of the singlet to be specified 
as well.  The pure singlet contribution $m_s^{2}\mu(\hat{S})$ in $\hat{W}$ 
is allowed by the symmetries of the model and the mass parameter $m_s^{2}$ 
is {\it a priori} arbitrary.  Through the $F$-term contributions, this 
induces a positive mass-squared parameter for the singlet: $m_{F}^{2}\equiv 
4 m_{s}^{4}/M_{Pl}^{2}$.  However, interactions at higher energies at or 
near the Planck scale can provide an additional mass-squared parameter 
$m_0^{2}$ as well as a quartic coupling $\lambda_s$.  Hence the effective 
potential for the singlet takes the form $V_s=M_s^2 |S|^{2}+\lambda_{s} 
|S|^{4}$ with $M_s^{2}\equiv m_0^2+m_{F}^{2}$. Since the Higgs doublets 
have vanishing $R$ charges, the electroweak breaking cannot have any effect 
on the fate of $U(1)_R$.  The only way to break it is to allow the singlet 
to develop a nonvanishing vacuum expectation value. This can happen only 
when $M_{s}^{2}<0$ so that $v_s^{2}= -M_s^2/2 \lambda_s$.  Since $m_F^2$ is 
positive, $m_0^2$ should be negative enough to induce a negative $M_s^2$. 
This impies that $m_s^2$ cannot be as large as $M_{Pl}^{2}$ as it would 
leave $U(1)_R$ unbroken; hence $|m_s^2|\sim |m_0^2|\sim v_s^2$ is a natural 
choice.  The singlet field could then be expanded around $v_s$ as
\begin{eqnarray}
\label{singlet}
S(x)= {1 \over \sqrt 2}[v_s + s(x)]~e^{i \varphi(x)},
\end{eqnarray}
where $\varphi(x)$ is the corresponding Nambu--Goldstone boson\cite{gold} 
which has a strictly flat potential, and $s(x)$ is a real scalar field with a 
mass of order $v_s$.  It is clear from the above that our $U(1)_R$ plays the 
role of what is usually called $U(1)_{PQ}$\cite{pq}.  Whereas the conventional 
$U(1)_{PQ}$ applies to the usual quarks and leptons, our $U(1)_{PQ}$ applies 
only to the superparticles, and the gluino is the only colored fermion 
having a nonvanishing $U(1)_{PQ}$ current:
\begin{eqnarray} 
J_{\mu}^{5,\tilde{g}}=\overline{\lambda^{a}}\gamma_{\mu}\gamma_{5} 
\lambda^{a}, 
\end{eqnarray}
where we have used the four-component notation: ${\lambda^{a}}=
(\tilde{\lambda}^{a}_{3}, \overline{\tilde{\lambda}^{a}_{3}})$. Now 
the gluino also contributes to the color current with respect to which 
$J_{\mu}^{5,\tilde{g}}$ has a nonvanishing quantum anomaly:
\begin{eqnarray}
\label{anomal}  
\partial^{\mu}J_{\mu}^{5,\tilde{g}}=\frac{6 g_s^2}{64 \pi^{2}} 
\epsilon_{\mu \nu \alpha \beta} G_a^{\mu \nu} G_a^{\alpha \beta}~.
\end{eqnarray}
Since $J_{\mu}^{5}$ couples to $\varphi$ as $\partial^{\mu}\varphi 
J_{\mu}^{5}$, the effective QCD vacuum angle takes the form 
\begin{eqnarray}
\overline{\theta}=\theta_{QCD} + 6 \varphi(x),
\end{eqnarray}
where the nondynamical phases in the quark mass matrices and the phase of the 
complex constant $k_3$ can be included in $\theta_{QCD}$ by a chiral rotation.
In close analogy with the KSVZ scenario\cite{ksvz}, our $\varphi(x)$ also 
receives a potential from the instanton background so as to develop a vacuum 
expectation value which enforces $\overline{\theta}\equiv 0$ [i.e. $\langle 
\varphi \rangle = -\theta_{QCD}/6$], to all orders in perturbation theory. 
Rather than the quarks, it is thus the gluino which realizes the Peccei-Quinn 
mechanism of solving the strong CP problem.

The axion, $a\equiv v_s [\varphi(x) - \left<\varphi\right>]$, has a mass and 
lifetime given by
\begin{eqnarray}
m_a\sim m_{\pi} \frac{f_{\pi}}{f_a}~, \ \ \ \tau(a\rightarrow 2 \gamma)\sim 
\left(\frac{m_{\pi}}{m_a}\right)^{5} \tau(\pi\rightarrow 2 \gamma)~,
\end{eqnarray}
where its decay constant $f_a$ is equal to $v_s/6$.  Our axion is not a 
DFSZ axion\cite{dfsz} as it does not couple to quarks and leptons; it is 
also not a KSVZ axion\cite{ksvz} as it does not couple to unknown colored 
fermion multiplets beyond the MSSM spectrum.  We may call it the 
$gluino$ $axion$\cite{prep} as it is induced by promoting the masses of the 
gauginos to local operators.

Let us choose $v_s/\sqrt 2 \sim 10^{11}$ GeV, which is in the middle of the 
range of $10^9$ to $10^{12}$ GeV allowed by astrophysical and cosmological 
bounds \cite{astro} on $f_a$. The effective theory below $v_s$ is then a 
replica of the MSSM with the effective $\mu$ parameter
\begin{eqnarray}
\label{effmu}
\mu_{eff}=\frac{v_s^{2}}{2 M_{Pl}}~e^{-i\theta_{QCD}/3}\sim 10^{3}~{\rm GeV}
\times e^{-i\theta_{QCD}/3}~,
\end{eqnarray}
which is the right scale for supersymmetry breaking.  This seesaw mechanism 
for the $\mu$ parameter results from the introduction of the composite 
operator given by Eq.~(\ref{composite}) into the theory, the dynamics of 
which are presumably dictated by physics at or near the Planck scale.  On the 
other hand, the scale of the $\mu$ parameter is fixed by the astrophysical 
and cosmological bounds on the axion decay constant $f_a$ which gives (or 
receives) a meaning to (from) the intermediate scale $v_s$ 
\footnote{If we abandon the composite operator idea (which correlates
the axion scale with $M_{SUSY}$), we can still get a phenomenologically 
acceptable model, though less economical than the present one, as follows:  
Let $S_2$ have $R=+2$ which couples as $\mu$.  Introduce $S_1$ with $R=+1$ 
which contains the axion.  Let $\langle S_2 \rangle \sim 1$ TeV, but 
$\langle S_1 \rangle \sim 10^{11}$ GeV.  The mixing between $S_1$ and $S_2$ 
is assumed small, so the axion coupling to the gluino is suppressed, i.e. 
a kind of DFSZ model applied to gluinos.}.
 
The low-energy effective theory is the softly broken MSSM with $R$ parity 
conservation. Indeed, after replacing the effective $\mu$ parameter 
[Eq.~(\ref{effmu})] and $\langle \varphi \rangle = -\theta_{QCD}/6$ into the 
effective Lagrangian [Eq.~(\ref{soft})], we obtain
\begin{eqnarray}
\label{softMSSM}
{\cal{L}}^{soft}_{MSSM}&=&\tilde{Q}^{\dagger} M_{Q}^{2} \tilde{Q} + 
\tilde{u^c}^{\dagger} M_{u^c}^{2} \tilde{u^c}+\tilde{d^c}^{\dagger}
M_{d^c}^{2}\tilde{d^c}+\tilde{L}^{\dagger} M_{L}^{2} \tilde{L}+
\tilde{e^c}^{\dagger} M_{e^c}^{2} \tilde{e^c}\nonumber\\&+&
\Big\{ A_{u} \tilde{Q}\cdot {H}_{u}~\tilde{u^c}+A_{d} \tilde{Q}\cdot 
{H}_{d} ~\tilde{d^c} + A_{e} \tilde{L}\cdot {H}_{d} ~\tilde{e^c}\big] + h. c.
\Big\}\nonumber \\ &+& M_{H_u}^{2} |H_u|^{2}+M_{H_d}^{2} |H_d|^{2}+
\left(\mu_{eff} B H_{u}\cdot H_{d} + h. c. \right)\nonumber\\
&+&\Big\{M_{3} \tilde{\lambda}^{a}_{3}\tilde{\lambda}^{a}_{3} +
M_2 \tilde{\lambda}^{i}_{2}\tilde{\lambda}^{i}_{2}+ M_1 \tilde{\lambda}_{1}
\tilde{\lambda}_{1}+ h. c. \Big\},
\end{eqnarray}
which is nothing but the soft supersymmetry-breaking part of the MSSM 
Lagrangian.  It is in fact this part of the Lagrangian that possesses all 
sources of CP violation through the complex $A$ parameters, the gaugino 
masses, and $\mu_{eff}$ itself. The explicit expressions for the mass parameters in 
Eq.~(\ref{softMSSM}) read as follows.  The gaugino masses are given by
\begin{eqnarray}
M_{3}=|k_3| \mu_{eff}^{*}~,\ \ \ M_{2}= k_2 \mu_{eff}^{*}~, \ \ \ M_{1}=k_1 
\mu_{eff}^{*}~, 
\end{eqnarray} 
which are not necessarily universal. The soft masses for the Higgs sector 
are given by
\begin{eqnarray}
M_{H_u}^{2}=y_u |\mu_{eff}^{2}|~,\ \ \ M_{H_d}^{2}=y_d |\mu_{eff}^{2}|~,\ \ \ 
\mu_{eff}B=|\mu_{eff}|^{2} (8 \frac{m_s^{2}}{v_{s}^{2}}+k_{\mu})~,
\end{eqnarray}
and are responsible for electroweak symmetry breaking, with similar 
expressions for the mass-squared matrices of the sfermion fields.  In 
particular, since $m_s^2/v_s^2$ is of order unity, the $B$ parameter is 
also of the same scale.  Finally the $A$ parameters are given by
\begin{eqnarray}
A_{u}=\mu_{eff}^{*}k_u~,\ A_d=\mu_{eff}^* k_d~,\ A_e=\mu_{eff}^* k_e~,
\end{eqnarray}
which do not have to be proportional to $h_u$, $h_d$, and $h_e$ of 
Eq.~(\ref{superpot}) as in the constrained MSSM.

As noted before, all mass scales of the MSSM Lagrangian [Eq.~(\ref{softMSSM})] 
are fixed in terms of $|\mu_{eff}|$.  More than this, the phase of 
$\mu_{eff}$, i.e. $-\theta_{QCD}/3$, contributes universally to all mass 
parameters which are complex.  However, the phases of the gaugino masses 
as well as those of the $A$ and $B$ terms also depend on the $k$ 
parameters.  Hence if the flavor structure of these matrices is not the 
same as those of the usual quarks and leptons, then the CP violation in 
flavor-changing processes is a powerful probe\cite{flavor} into this 
sector of the effective theory.  In the calculation of electric dipole 
moments due to supersymmetry\cite{susycp}, these CP phases can be considered 
as they are without worrying about whether there is an additional contribution 
from $\overline \theta$.

In conclusion, we have presented in the above a simultaneous solution to 
two hierarchy problems, i.e. why $\overline \theta$ is so small (the strong 
CP problem) and why $\mu$ is 1 TeV and not $10^{16}$ GeV (the $\mu$ problem), 
as well as the related issue of why $M_W$ and $M_{SUSY}$ are only one order 
of magnitude apart.  The primary difference between our approach and 
previous other attempts\cite{seek} lies in the fact that the gaugino masses 
are promoted here to local operators given by $\mu(\hat{S})$. Indeed, finite 
bare mass terms for the gauginos would automatically break the $U(1)_{PQ}$ 
symmetry, making it impossible for the relaxation of $\overline{\theta}$ to 
zero.  As it is, $\langle S \rangle$ serves two important purposes: its 
magnitude determines the scale of supersymmetry breaking and its phase 
solves the strong CP problem.

Let us summarize our proposal.
 
($i$) We work in the framework of supersymmetry and identify $U(1)_{PQ}$ 
as $U(1)_R$ which contains the usual $R$ parity as a discrete subgroup.

($ii$) We require the supersymmetric Lagrangian and all supersymmetry 
breaking terms to be invariant under $U(1)_R$.

($iii$) We implement this with the composite operator $\mu(\hat S) \equiv 
(\hat S)^2/M_{Pl}$ where $\hat S$ is a singlet superfield having $R = +1$.

($iv$) The spontaneous breaking of $U(1)_R$ generates an axion and relaxes 
the effective QCD vacuum angle $\overline \theta$ to zero, using the 
dynamical gluino phase, thus solving the strong CP problem.

($v$) The existing astrophysical and cosmological bounds on the axion decay 
constant implies a supersymmetry breaking scale of 1 TeV.

($vi$) The effective Lagrangian at low energy is that of the MSSM with $R$ 
parity conservation.  All mass scales are of order 1 TeV, thus solving 
the $\mu$ problem and the related issue of why $M_W$ and $M_{SUSY}$ are only 
an order of magnitude apart.

($vii$) Since $\overline \theta = 0$ in this consistent supersymmetric 
theory, electric dipole moments can be calculated unambiguously from the 
other explicit CP violating terms of the MSSM.

\newpage
D.A.D. acknowledges the hospitality of the UCR Physics Department, where 
this work was initiated.  The research of E.M. was supported in part by the 
U.~S.~Department of Energy under Grant No.~DE-FG03-94ER40837.


\begin{thebibliography}{99}
\bibitem{exp1} J. H. Christenson {\it et. al.}, Phys. Rev. Lett. {\bf 13}, 
138 (1964).
\bibitem{exp2} A. Alavi-Harati {\it et al.}, Phys. Rev. Lett. {\bf 83}, 22 
(1999)
\bibitem{exp3} P. G. Harris {\it et. al.}, Phys. Rev. Lett. {\bf 82}, 904 
(1999).
\bibitem{theta} C. G. Callan, R. Dashen, and D. J. Gross, Phys. Lett. 
{\bf 63}, 334 (1976); R. Jackiw and C. Rebbi, Phys. Rev. Lett. {\bf 37}, 
172 (1976).
\bibitem{edm} V. Baluni, Phys. Rev. {\bf D19}, 2227 (1979); R. Crewther, P. 
DiVecchia, G. Veneziano, and E. Witten, Phys. Lett. {\bf 88B}, 123 (1979).
\bibitem{pq} R. Peccei and H. Quinn, Phys. Rev. Lett. {\bf 38}, 1440 (1977).
\bibitem{ww} S. Weinberg, Phys. Rev. Lett. {\bf 40}, 223 (1978); 
F. Wilczek, Phys. Rev. Lett. {\bf 40}, 279 (1978).
\bibitem{cancel} T. Ibrahim and P. Nath, Phys. Rev. {\bf D57}, 478 (1998);
Phys. Lett. {\bf B418}, 98 (1998); Phys. Rev. {\bf D58}, 111301 (1998);
hep-ph/9910553; M. Brhlik, G. J. Good, and G. L. Kane, Phys. Rev. {\bf D59}, 
115004 (1999).
\bibitem{susycp} M. Dugan, B. Grinstein, and L. J. Hall, Nucl. Phys. 
{\bf B255}, 413 (1985); E. Ma and D. Ng, Phys. Rev. Lett. {\bf 65}, 2499 
(1990); D. A. Demir, Phys. Rev. {\bf D60}, 055006, 095007 (1999); A. 
Pilaftsis and C. E. Wagner, Nucl. Phys. {\bf B553}, 3 (1999).
\bibitem{exp4} L. J. Rosenberg and K. A. van Bibber, Phys. Rep. {\bf 325}, 1 (2000).
\bibitem{dfsz} M. Dine, W. Fischler, and M. Srednicki, Phys. Lett. {\bf B104},
199 (1981); A. Zhitnitskii, Sov. J. Nucl. Phys. {\bf 31}, 260 (1980).
\bibitem{ksvz} J. E. Kim, Phys. Rev. Lett. {\bf 43}, 103 (1979); M. A. 
Shifman, A. Vainshtein, and V. Zakharov, Nucl. Phys. {\bf B166}, 493 (1980).
\bibitem{dith} S. Dimopoulos and S. Thomas, Nucl. Phys. {\bf B465}, 23 (1996);
D. A. Demir, hep-ph/9911435. 
\bibitem{transf} Under $U(1)_R$, the scalar components of a chiral 
superfield transforms as $\phi \to e^{i \theta R} \phi$, whereas the 
fermionic components transform as $\psi \to e^{i \theta (R-1)} \psi$.
\bibitem{gold} Y. Nambu, Phys. Rev. Lett. {\bf 4}, 380 (1960); J. Goldstone, 
Nuovo Cimento {\bf 19}, 154 (1961).
\bibitem{prep} Details such as the precise $a \to \gamma \gamma$ coupling 
and other phenomenological implications will be discussed in a forthcoming 
paper: D. A. Demir and E. Ma, in preparation.
\bibitem{astro} G. G. Raffelt, Ann. Rev. Nucl. Part. Sci. {\bf 49}, 163 
(1999).
\bibitem{flavor} D. A. Demir, A. Masiero, and O. Vives, Phys. Rev. {\bf D61}, 
075009 (2000); hep-ph/9911337.
\bibitem{seek} J. E. Kim and H. P. Nilles, Phys. Lett. {\bf 138B}, 150 
(1984);  W. Buchm\"uller and D. Wyler, Phys. Lett. {\bf 121B}, 321 (1983); 
J. E. Kim, Phys. Rep. {\bf 150}, 1 (1987).
\end{thebibliography}
\end{document}